\documentclass[prl,notitlepage, twocolumn]{revtex4-1}
\usepackage{amsmath,mathtools,amssymb}
\usepackage{braket,bm,bbold}
\usepackage{physics}
\usepackage{empheq}
\usepackage{simpler-wick}

\renewcommand\theequation{\thesection\arabic{equation}}

\graphicspath{{images/}}
\usepackage[%
colorlinks=true,
urlcolor=blue,
linkcolor=blue,
citecolor=blue
]{hyperref}
   
\newcommand{\Kp}{{\bar{K}}}

\begin{document}
\title{Excitonic fractional quantum  Hall hierarchy in Moir\'e heterostructures}
\date{\today}
\author{Yves H. Kwan}
\affiliation{Rudolf Peierls Centre for Theoretical Physics,  Clarendon Laboratory, Oxford OX1 3PU, UK}
\author{Yichen Hu}
\affiliation{Rudolf Peierls Centre for Theoretical Physics,  Clarendon Laboratory, Oxford OX1 3PU, UK}
\author{Steven H. Simon}
\affiliation{Rudolf Peierls Centre for Theoretical Physics,  Clarendon Laboratory, Oxford OX1 3PU, UK}
\author{S. A. Parameswaran}
\affiliation{Rudolf Peierls Centre for Theoretical Physics,  Clarendon Laboratory, Oxford OX1 3PU, UK}
\begin{abstract}
We consider  fractional  quantum Hall states in systems where two flat Chern number $C=\pm 1$ bands are labeled by an approximately conserved  `valley' index and interchanged by time reversal symmetry. At filling factor $\nu=1$ this setting admits an unusual hierarchy of  correlated phases of {\it excitons}, neutral particle-hole pair excitations of a fully valley-polarized `orbital ferromagnet' parent state where all electrons occupy a single valley. Excitons experience an effective magnetic field due to the Chern numbers of the underlying bands. This obstructs their condensation in favor of a variety of crystalline orders and gapped and gapless liquid states. All these have the {\it same} quantized charge Hall response and are electrically incompressible, but differ in their edge structure, orbital magnetization, and hence valley and thermal responses. We explore the relevance of this scenario for Moir\'e heterostructures of bilayer graphene on a hexagonal boron nitride substrate.
\end{abstract}
\maketitle
The observation of gate-tunable superconductivity and  correlated insulating behaviour  in twisted bilayer graphene (TBG) in the magic angle regime~\cite{cao2018a,cao2018b}  has  stimulated intense  investigation of  two-dimensional (2D) van der Waals heterostructures. 
 While the precise mechanism behind the  high superconducting transition temperatures (relative to the low carrier density) remains hotly debated,
intrinsic to this setting is the enhancement of correlations when the electronic dispersion is reconstructed by the interlayer Moir\'e pattern corresponding to a small twist angle.  After accounting for spin and valley degeneracies, substrate-free TBG hosts eight flattened central bands that are interlinked by Dirac points and energetically separated from remote bands. This quenching of the single-particle kinetic energy bears a family resemblance to the formation of Landau levels (LLs) by 2D electron gases in magnetic fields. The analogy is sharpened in the presence of an encapsulating hexagonal boron nitride (hBN) substrate: this opens topological gaps at the  Dirac points, pushing four bands  above (below) the neutrality point while assigning each band a non-zero Chern number ($C=\pm1$) in a manner that preserves overall time-reversal symmetry (TRS)~\cite{bultinck2019,zhang2019a,zhang2019b}.

Flat Chern bands  with $|C|=1$  are similar to LLs: when fully filled, they show a quantized anomalous Hall (QAH) response~\cite{TKNN,haldane1988,liu2016}, the lattice analogue of the integer quantum Hall  (QH) effect~\cite{PrangeGirvin}. At  commensurate partial fillings, interactions can stabilize  incompressible fractional Chern insulators (CIs)~\cite{parameswaran2013,bergholtz2013}. However, most realizations of CIs, e.g. in magnetic topological insulators~\cite{chang2013} or cold atomic gases~\cite{jotzu2014} have significant single-particle dispersion and hence relatively weak correlations. 

The observation of a QAH response in hBN-TBG devices at filling $\nu=+3$ relative to charge neutrality in the absence of an external magnetic field~\cite{serlin2019} points to the breaking of TRS by interactions, leading to selection of a spin- and valley-polarized insulating state corresponding to fully filling a single Chern band~\cite{bultinck2019,xie2018,zhang2019b}. While reminiscent of   quantum Hall ferromagnetism (QHFM) in LLs, an important distinction in hBN-TBG is that TRS is broken {\it spontaneously}, lifting the degeneracy between two valleys with equal and opposite Chern number.  These systems are  robust and tunable platforms to study correlated Chern insulators and proximate phases~\cite{repellin2019chern,ledwith2019fractional}.

Here, we propose  that systems such as hBN-TBG, where two nearly-flat degenerate Chern bands have equal and opposite Chern number, are unique settings for an unconventional {\it excitonic quantum Hall hierarchy}.  The excitons we consider are stable neutral gapped intervalley excitations of the fully-valley-polarized insulator, that are bound states of a hole in a filled Chern band and a particle in an empty band with the opposite Chern number.  Tightly-bound excitons can be viewed --- in a sense we make precise in a companion paper~\cite{KHSP2020a}
 --- as neutral bosons in a Chern band. As in TBG we take the bands to correspond to different  valleys; assuming valley conservation we may then meaningfully view the excitons as `filling' a Chern band. The exciton filling tracks  the change in valley polarization relative to the fully-polarized parent state. We show using a simplified LL model~\cite{bultinck2019} that  the fully-valley polarized phase is proximate to a rich hierarchy of correlated phases that emerge when interactions between excitons lead them to form  incompressible bosonic FQH liquid states, a variety of Wigner crystal or stripe/bubble phases with broken translational symmetry, or compressible Fermi-liquid-like states (CFLs).  All the incompressible states  (and some of the compressible ones) have {identical}  charge response, namely a quantized Hall conductivity $\sigma_{xy} =  e^2/h$ and vanishing longitudinal conductivity $\sigma_{xx}=0$, but differ in their valley and thermal Hall responses.
 We suggest  experimental probes   to distinguish the various excitonic phases. Finally, we discuss how the delicate balance of energy scales from interactions, gate screening, and  the  residual band dispersion could stabilize this excitonic FQH hierarchy within the phase diagram of hBN-TBG or other Moir\'e systems.

\textit{Model and Fully Valley-Polarized Parent State.---} We are ultimately interested in  the eight central bands of TBG at fillings $\nu = \pm 3, \pm1$. Here, $\nu=0$ corresponds to charge neutrality, and $\nu=-4(+4)$ corresponds to the case where all these eight bands  are empty (filled) [Fig.~\ref{fig:bandsVexc}(a)].  The hBN substrate opens a single-particle gap at neutrality resulting in bands with Chern numbers $C^{>}_{K,\sigma}=-C^{<}_{K,\sigma} = - C^{>}_{\Kp,\sigma}=C^{<}_{\Kp,\sigma}=1$
where  $>$ ($<$) labels bands above (below) neutrality and we have introduced valley $\tau = K, \Kp$ and   spin $\sigma=\uparrow, \downarrow$ labels~\cite{bultinck2019,xie2018,zhang2019b}.  Throughout, we ignore spin-orbit coupling and elevate  approximate  valley conservation to an exact $U(1)_v$ symmetry. The non-interacting band structure thus has   $SU(2)_s\times U(1)_v\times U(1)_c$ symmetry (where $c, v ,s$ refer to charge, valley, and spin), and preserves TRS which interchanges the valleys and flips the sign of $C$.

In order to study interaction effects at odd integer  filling we introduce several simplifications. First, we work with a model interaction projected to the relevant bands and ignore mixing between bands split at the single-particle level. Second, we suppress the  spin degree of freedom and restrict our attention to the partially filled doublet of  degenerate   valleys 
$K, \Kp$ with equal and opposite Chern numbers (Fig.~\ref{fig:bandsVexc}a).  Finally, in line with previous studies~\cite{bultinck2019} we replace the Chern bands  with LLs where valleys $K, \Kp$ see equal and opposite magnetic fields. We note that these are reasonable approximations appropriate to the flat-band limit of interest that nevertheless capture the underlying topological  band structure. The  single-particle Hamiltonian in   valley $\tau=\pm$ (henceforth we use valley index and Chern number interchangeably as they are tied together in the two-valley subspace) takes the form 
$H_{\pm} =  \frac{\left(\boldsymbol{p} \mp e \boldsymbol{A}\right)^2}{2m}$ with $\boldsymbol{\nabla}\!\times\!\boldsymbol{A} = B$. 
We assume each valley is in its $N_\Phi$-fold degenerate lowest LL; here  $N_\Phi = A/2\pi \ell_B^2$ counts the number of flux quanta threading sample  area $A$, and  $\ell_B = (\hbar /eB)^{1/2}$ is the magnetic length (which plays the role of the Moir\'e scale in TBG). We fix  the  filling factor of this pair of LLs at $\nu =  \nu_+ + \nu_{-} = 1$, where $\nu_\pm = N_\pm/N_\Phi$  is the filling factor in each valley. The  effective Hamiltonian consists of interactions projected onto the degenerate LLs,
\begin{equation}\label{eq:mainHam}
H_{\text{int}} = \frac{1}{2N_\Phi} \sum_{\mathbf{q},{\tau,\tau'}} V_{\tau\tau'}(\mathbf{q}) :\bar{\rho}_\tau(\mathbf{q}) \bar{\rho}_{\tau'}(-\mathbf{q}) :.
\end{equation}
 Here, we have introduced the projected density operators 
\begin{equation}
\bar{\rho}_\pm (\mathbf{q})  = F(\mathbf{q}) \sum_{k_y} e^{\pm i q_x k_y \ell_B^2} c^\dagger_{k_y - \frac{q_y}{2}, \pm}c^{\phantom\dagger}_{k_y + \frac{q_y}{2},\pm},
\end{equation}
where $F(\mathbf{q}) = e^{-\mathbf{q}^2\ell_B^2/4}$ and $c^\dagger_{{k}_y,\pm}$ is the creation operator for a Landau-gauge single-particle lowest-LL state $\phi_{k_y,\pm}(x)  = \frac{1}{\sqrt{L_y \ell_B \sqrt{\pi}}}e^{i k_y y} e^{-{(x \mp k_y \ell_B^2)^2}/{2\ell_B^2}}.$
We  choose a phenomenological interaction that only includes density-density  terms,
$V_{\tau,\tau}(\mathbf{q}) = v(\mathbf{q}) =\frac{2\pi e^2}{|\mathbf{q}|}$, $V_{\tau,-\tau}(\mathbf{q}) = v_d(\mathbf{q}) =v(\mathbf{q}) e^{-|\mathbf{q}|d}$,
where $d$  tunes the competition of inter- and intra-valley interactions. {Other intervalley terms  are  $o(a_0/\ell_B)$  where $a_0$ is a  lattice scale linked to the separation  of valleys in the microscopic Brillouin zone (BZ).   

Eq.~\eqref{eq:mainHam} is essentially the LL limit of a minimal model for TBG introduced in Ref.~\cite{bultinck2019}, absent a periodic potential and with  slightly modified interactions. There, within a Hartree-Fock (HF) analysis it was argued that the ground state of ~\eqref{eq:mainHam} is a  fully-valley-polarized  insulator (FVPI) that with $\nu_+ =1, \nu_{-}  =0$. This state has a QAH response linked to the spontaneous breaking of TRS. The FVPI was argued to be stable against both the inclusion of a weak nonzero single-particle dispersion, as well as allowing  
$d>0$. Although the former is also true for  conventional QHFMs where all bands have $C=1$, the latter is unexpected:  arguing in analogy with bilayer QH systems, we would anticipate that   `softening' the intervalley interactions in this way would stabilize intervalley-coherent states with $\nu_+ = \nu_{-}=\frac{1}{2}$. However when $C_+ = -C_-$, even for 
$d\rightarrow 0$  the symmetry is reduced relative to the  $C_+ = C_-$~\cite{bultinck2019} case. This is evident, e.g., in the gap to ``valley-flip'' excitations of the FVPI, that persists in more microscopic models of TBG~\cite{KHSP2020a,wu2020}.
 
\textit{Exciton Topology and Interactions.---} A more striking consequence of the reversal of Chern numbers between the valleys lies in the topological structure of intervalley excitations. Consider a single inter-valley particle-hole pair excitation of the FVPI. Up to an overall constant loss of exchange energy in creating a single hole, in the LL limit the Hamiltonian is that of an electron and a hole in equal and opposite magnetic fields $\mp B$, with {\it attractive} inter-valley interactions, 
$H_{\text{ex}} =  \frac{\left(\boldsymbol{p}_e + e \boldsymbol{A}\right)^2}{2m} + \frac{\left(\boldsymbol{p}_h + e \boldsymbol{A}\right)^2}{2m} - v_d(\mathbf{r})$,  where $v_d(r) = \int \frac{d\mathbf{q}}{(2\pi^2)} e^{-i\mathbf{q}\cdot \mathbf{r}} v_d(\mathbf{q})$.
This decouples in  terms of relative $\mathbf{r} = \mathbf{r}_h -\mathbf{r}_e$ and center-of-mass $\mathbf{R} = \frac{\mathbf{r}_h +\mathbf{r}_e}{2}$ coordinates, yielding $H_{\text{ex}} 
 = H_{\mathbf{R}}+H_{\mathbf{r}}$, with
 \begin{equation}\label{eq:CMrel}
H_{\mathbf{R}} =   \frac{\left(\boldsymbol{P}_{\text{CM}}+ 2 e \boldsymbol{A}\right)^2}{4m}, H_{\mathbf{r}} =  \frac{\left(\boldsymbol{p}_{\text{rel}}+  \frac{e}{2}\boldsymbol{A}\right)^2}{m} - v_d(r).
\end{equation}
Accordingly, each discrete excitonic bound-state solution of $H_{\mathbf{r}}$ has a $2N_\Phi$-fold degeneracy corresponding to the LL degeneracy of $H_{\mathbf{R}}$. 
Explicitly, if $z_e, z_h$  are complex coordinates for the electron and hole respectively, 
 defining $z = \frac{z^e+z^h}{2}$ and $u= z^h-z^e$ and freezing the relative coordinate $u$ in the lowest excitonic  bound state $\phi_0$ yields
\begin{equation}\label{eq:excproj}
\psi_{\text{exc}}(z, u) \!=\!  \tilde{f}(z) e^{-|z|^2/2} \phi_{0}(u)\text{ with }\phi_0(u) \!\equiv\! e^{-|u|^2/8},\!\!\!\!
\end{equation}
where $\tilde{f}$ is analytic and we take $\ell_B=1$~\cite{SupMat}. Note that $z$ ($u$) sees an effectively doubled (halved)  field, consistent with \eqref{eq:CMrel}. Exciton structure  is  more complicated in realistic models that include the dispersion  and Berry curvature fluctuations of the underlying bands, and both the form of the `envelope function' $\phi_0$ and its coupling to center-of-mass motion can influence exciton topology. Nevertheless, the lowest exciton band has $C\neq 0$ for a range of parameters even in realistic models of hBN-TBG~\cite{KHSP2020a}. Even beyond this regime,  exciton  bands have substantial Berry curvature, which can influence phase structure even if the Chern number (its integral over the BZ of allowed exciton momenta~\cite{KHSP2020a}) vanishes. 

Each exciton also carries a unit of  $U(1)_v$ valley polarization. Since the latter is conserved it is  meaningful to consider partial valley-polarized states corresponding to a finite density of excitons. For instance, intervalley-coherent `exciton condensate'  HF trial states with $\nu_+=\nu_- =\frac{1}{2}$  are energetically uncompetitive in much of the phase diagram because exciton topology forces them to host a vortex lattice analogous to Type-II superconductors in a magnetic field~\cite{bultinck2019}. Evidently, as bosons in a magnetic field the excitons can form various other many-body states inaccessible to a HF analysis, depending on the effective exciton-exciton  interaction $V_{\text{exc}}(R)$. For our choice of $v(\mathbf{r})$  we estimate $V_{\text{exc}}(R)$ by considering appropriately antisymmetrized variational wavefunctions for a pair of tightly-bound excitons at separation $R$~\cite{SupMat}; representative results are sketched in Fig.~\ref{fig:bandsVexc}(b).  Short-range interactions are repulsive (attractive) if $d/\ell_B \gtrsim1$ ($\lesssim$).  For any $d>0$ excitons experience an asymptotic $R^{-3}$ repulsion that vanishes for $d=0$. (A power-law tail is generically expected, but its exponent may depend on the choice of $v(\mathbf{r})$.) A second subtlety  in considering many-body exciton phases is that single-exciton states are not all independent~\cite{yangexciton}; however, this is unimportant in the dilute limit  $\nu_- = 1-\nu_+ \ll 1$ and we ignore this below.

\textit{Exciton FQH Hierarchy.---} With these preliminaries, we now turn to possible excitonic phases. We consider fillings $(\nu_+, \nu_-) = (1-\nu_v, \nu_v)$ with $0<\nu_v<\frac{1}{2}$, corresponding to adding $\nu_vN_\Phi$ intervalley excitons to the FVPI resulting in  valley polarization $I^z_v = \frac{1}{2}-\nu_v$ per electron. 
 Since all terms have characteristic scale $\sim e^2/\ell_B$ the  phase structure  turns on microscopic details of the interactions. For now we  assume that  exciton binding sets the dominant scale and that the inter-exciton interaction $V_{\text{exc}}$  is always repulsive and short-ranged. (We comment on other cases below.) In this regime, it is reasonable to assume  that the  exciton  is a stable bound state. 
Since each exciton behaves as though it occupies a $2N_\Phi$-fold degenerate LL, the effective  filling factor is $\nu_b = \nu_v/2$. For  $\nu_v = 1/m$ with $m$ an integer, the tightly-bound excitons can form a $\nu_{b} = 1/2m$ bosonic Laughlin state. A trial wavefunction that captures  this is given by 
\begin{eqnarray}
\Psi_{2m}(\{z^e,z^h\}) &=& \mathcal{P}_{\text{exc}}[\Psi_m(\{z^e\})\Psi_m(\{z^h\})] \label{eq:excFQHwf}  
\\
&=& \sum_{\sigma\in S_N} {\text{sgn}\, \sigma}[\Psi_{m}(\{z_{i,\sigma(i)}\})^2\prod_j \phi_0(u_{j\sigma(j)})].\nonumber
\end{eqnarray}
Here, $\Psi_m(\{z\}) = \prod_{i<j}(z_i-z_j)^m e^{-\sum_i \frac{|z_i|^2}{4}}$ and $\mathcal{P}_{\text{exc}}$ projects electron-hole pairs into the excitonic ground state $\phi_0$~\eqref{eq:excproj}. This antisymmetrizes over permutations $\sigma\in S_N$ corresponding to different `pairings'  of the $N = N_\Phi/m$ electrons $i$ and holes $\sigma(i)$ to form excitons centered at $z_{i \sigma(i)} = \frac{1}{2}(z^e_i + z^h_{\sigma(i)})$ at separation $u_{i \sigma(i)} = z^h_i - z^e_{\sigma(i)}$ and then projects the latter into $\phi_0$ ~\cite{SupMat}. 
$\Psi_{2m}(\{z^e,z^h\})$ is a many-particle state of electrons in valley `$-$'  and holes in valley  `$+$', built on top of the FVPI parent state `vacuum'. Particle-hole (PH) transforming  \eqref{eq:excFQHwf}  only in the `$+$' valley  yields a purely electronic wave function. 

An alternative picture of the excitonic phase structure is obtained by viewing the problem (after PH transformation to holes in valley `$+$') as a two-valley system where each component sees the {\it same}  magnetic field and is at filling $\nu = 1/m$, and where inter (intra) valley interactions are {attractive} (repulsive)~\cite{zhang2018composite}. This gives distinct pictures for odd and even $m$. (In each case, the electronic state is obtained after undoing the PH transformation.)
\begin{figure}
\includegraphics[width=\columnwidth,trim={0.2cm 22.8cm 5.2cm 0.0cm},clip=true]{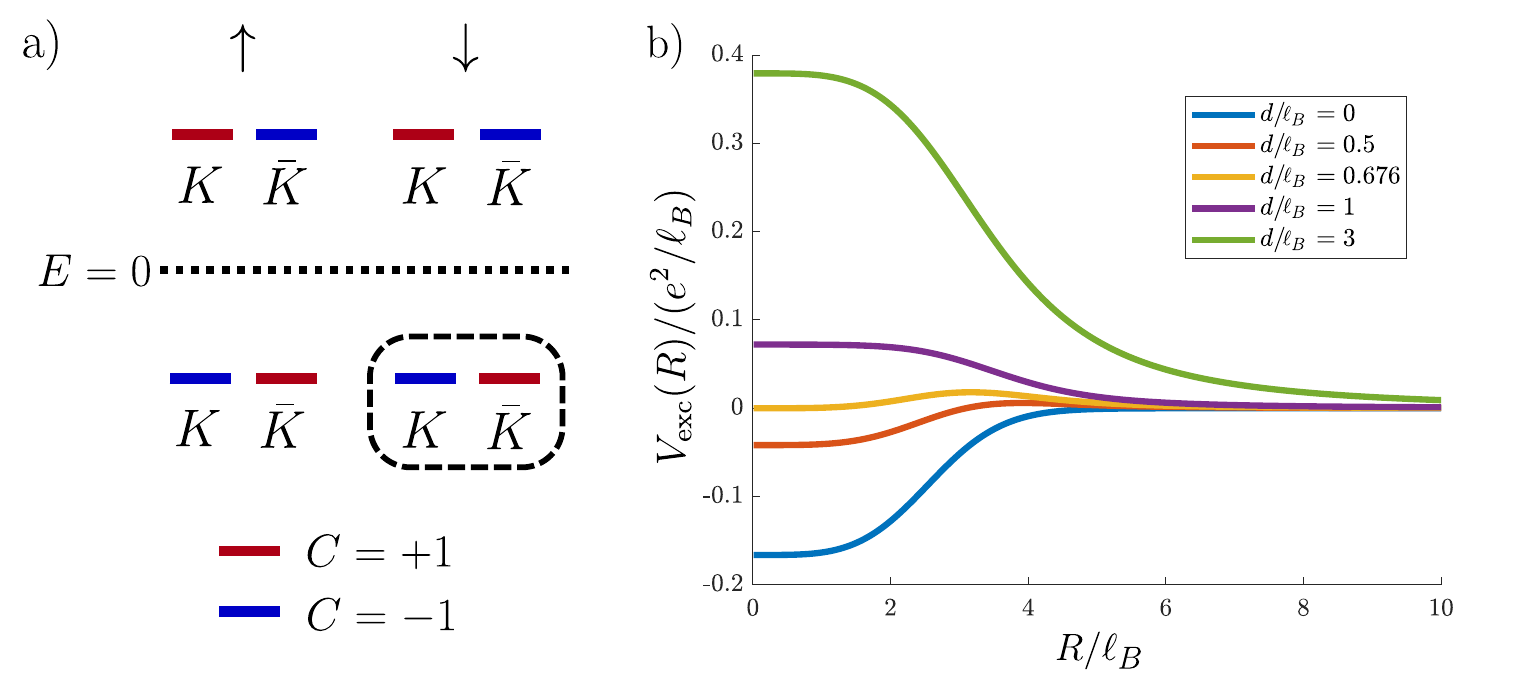}
\caption{\label{fig:bandsVexc}(a) Schematic of eight central  bands and Chern numbers in hBN-TBG; our model focuses on a pair of these (dashed box). (b) Exciton-exciton interaction profile as competition of inter/intra-valley interactions is tuned by $d$.}
\end{figure}

For even $m$, we can attach $m$  quanta of flux to each valley separately, yielding an equal density of composite fermions (CFs) in each valley.
For weak intervalley attraction we anticipate that these are stable against pairing~\cite{CFLCooper}, yielding a compressible state. For increasing attraction, we expect  a transition into an inter-valley paired state of CFs, schematically given by 
\begin{equation}\label{eq:CFWF}
\Psi^\text{CF}_{2m}\sim\mathcal{P}_L \prod_{i<j}(z^e_i-z^e_j)^m \prod_{k<l}(z^h_k-z^h_l)^m\text{det}[g(z^e_i - z^h_j)]
\end{equation}
where the determinant describes pairing with wavefunction $g(z)$, $\mathcal{P}_{L}$ projects to the lowest LL and we have omitted Gaussian factors. For $s$-wave or strong-coupling higher-angular momentum pairing we expect $g(z) \sim e^{-z/\xi}$ as $z\rightarrow\infty$, where $\xi\sim o(\ell_B)$ is  the pair size. Qualitatively, this pairs electrons and holes into tightly-bound bosons that then form a $\nu_b=1/2m$ Laughlin state. This is consistent with our picture of \eqref{eq:excFQHwf}, so we conclude that the two approaches describe similar physics. In non-$s$-wave cases, strong- and weak-pairing regimes are separated by a phase transition. For $p_x+ip_y$-pairing where $g(z) \sim 1/z$, this would be a transition between state \eqref{eq:excFQHwf} and the $(m-1,m-1,1)$ Halperin state which also has $\nu=1/m$ in each valley. (The equivalence of \eqref{eq:CFWF} with $p_x+ip_y$ pairing to the Halperin state follows via the Cauchy identity~\cite{MoonCoherence,BilayerPairedQHDrag}.) For $m=4$, preliminary exact diagonalization studies~\cite{SupMat} find a unique ground state for $N=4,6,8,10$ particles on the sphere~\cite{HaldaneSphere} at a shift~\cite{wenzeeshift} appropriate to \eqref{eq:excFQHwf} for certain short-range interactions, suggesting it is energetically competitive (since there is no other obvious incompressible candidate at this shift).

For $m$ odd and zero inter-valley interactions each component forms an {\it independent} $\nu=1/m$ fermionic Laughlin state. Inter-valley attraction  `locks' these together, suppressing  fluctuations where a particle in one valley is far from a hole in the other.   However, as each valley is independently incompressible,  numerical observation of a unique ground state is less strong evidence for  \eqref{eq:excFQHwf}.

Other unconventional phases are also possible. For  example, if $\nu_v = 2/q$ with $q$  odd,  the exciton filling is $\nu_b = 1/q$ ruling out bosonic Laughlin states. In this limit,  attaching $q$ quanta of $U(1)_v$ valley flux  to each exciton gives rise to a compressible  excitonic composite Fermi liquid (e-CFL). This does not obviously decouple into separate flux attachments to constituent electrons/holes. The e-CFL has inter-valley binding but no valley coherence, and is hence distinct from interlayer coherent CFLs~\cite{aliceainterlayerCFL} proposed  in QH bilayers. A more exotic  possibility is that the e-CFL in turn can undergo $p$-wave pairing to form a non-Abelian QH state.  The states considered here are specific examples of a rich hierarchy of FQH states of excitons, whose detailed analysis we defer to future work.

\textit{Edge Structure and Response.---} We now  discuss transport properties and  bulk response  of the states considered above. Since the excitonic Laughlin state \eqref{eq:excFQHwf} is a bosonic FQH state, we expect a quantized response in the `charge' carried by this state, which (translating back into the underlying electrons) leads to a  fractional quantized valley Hall (QVH) response, $\sigma^v_{xx}=0$, $\sigma^v_{xy}= -\nu_b\frac{q_v^2}{h}= -\frac{1}{2m}\frac{q_v^2}{h}$, where  $q_v=1$  is the  `valley charge' of a  single exciton. We can understand this also from an  edge-state perspective. Before implementing exciton projection, in terms of the underlying electrons we can view the edge of \eqref{eq:excFQHwf}  as built out of (i) a chiral $\nu_+=1$ chiral mode  of electrons  in a  filled LL;  (ii) a $\nu_+= 1/m$ chiral edge mode of  holes in valley `$+$'; and  (iii) a chiral $\nu_-=1/m$ edge of electrons in valley `$-$'. Owing to the opposite charge of holes and the opposite sign of $B$ in the  two valleys (ii) and (iii)  counter-propagate relative to  (i) and co-propagate  relative to each other. The  exciton projection can then  be viewed as  binding the  two co-propagating   fermionic  FQH edge  modes due to the attractive electron-hole interactions, leading to a single bosonic  $\nu_b  =  1/2m$ mode propagating `upstream' of the charge mode.

Other cases are  more complicated. For $\nu_v =  1/m$ with $m$ even and  weak inter-valley interactions, we find decoupled CFL-like states~\cite{zhang2018composite} in which the charge QH response breaks down in favor of metallic transport. In contrast, for $\nu_v = 2/q$ with $q$ odd, since the exciton binding dominates, both compressible and incompressible phases have a charge QH response. In the compressible e-CFL state this coexists with `metallic' valley response from  the exciton Fermi surface  whereas the incompressible paired descendants of the e-CFL  have a QVH response.  

Experimentally, it is challenging to distinguish different exciton phases via electrical measurements, since nearly all of them have identical $\sigma_{xy}  =e^2/h$  charge QH response. While it is difficult to directly measure the QVH response, `upstream' modes can be detected by measuring thermal  conductance $K_H$~\cite{KaneFisherThermal}. If  the upstream and downstream  modes are fully thermally equilibrated,  $K_H=0$,  whereas if they are out of equilibrium we expect a {\it doubled} response  relative to the integer QH case. Phases with QVH response will also show plateaus in orbital magnetization  quantized at a rational fraction of its value  in the FVPI, that could be optically detected.

\textit{Crystalline Phases.---} So far we have ignored the long-range tail of $V_{\text{exc}}(R)$. While this is unlikely to destabilize FQH liquids favored by short-range repulsive interactions, the competition between short-range attraction and long-range repulsion~\cite{Spivelson} can drive the formation of bubble and stripe phases~\cite{Fogler2001}. Similarly, at higher density e.g. $\nu_v =\frac{1}{2}$ Wigner-crystal like phases of excitons (analogous to the exciton vortex lattice) can be formed. These states, whose study we defer to the future, all  have broken translational symmetry, and (if pinned by the Moir\'e potential or disorder) can also show a charge QH response.

\textit{Discussion.---} We  have proposed  a new class of  FQH state formed by the  binding of electron-hole pairs in bands with equal and opposite Chern number. In closing, we return to our original goal of linking this to the  physics of TBG, which has several key ingredients ---  flipped Chern numbers,  flat bands,  and interactions --- that were pertinent to our analysis. However, our model leaves out other features such as band dispersion  and  Berry curvature fluctuations. Another concern is that numerics indicate that incompressible excitonic states are more stable if intravalley interactions are larger than intervalley couplings, whereas for TBG they may be comparable. 
Since the valley-dependence of the short-range component of interactions is difficult to precisely determine, it is reasonable to explore a wider parameter regime allowing for asymmetry of intra-and inter-valley couplings.  HF studies~\cite{bultinck2019}  in a similar regime  indicate that their competition combined with  nonzero  dispersion  can stabilize partially valley-polarized metals against the FVPI. It seems possible that an {\it incompressible} partially-valley polarized insulating excitonic FQH state may be energetically competitive to these. The tuning of valley occupation by magnetic field may also  provide a route to stabilizing excitonic phases. Another notable omission is any discussion of spin, whose influence on inter-exciton interactions may further enrich the phase  diagram.  Beyond TBG, multilayer Moir\'e heterostructures may have capacitive charging effects favoring excitonic states, and their valley and layer degrees of freedom may be intertwined so that the QVH response becomes accessible. In the future, it will be interesting to explore the relevance of  these ideas to other Moir\'e and flat-band systems or strained graphene~\cite{GhaemiStrain} and clarify their connection to  `fractional  excitonic insulators' proposed to form near topological band inversions far from the flat band limit~\cite{FracExcIns}.

\textit{Acknowledgements.---} We thank N. Bultinck,  B. Lian, N.~Regnault, S.L. Sondhi, and M.P. Zaletel for useful discussions.
We acknowledge support from support from the European Research Council (ERC) under the European Union Horizon 2020 Research and Innovation Programme (Grant Agreement No.~804213-TMCS) and from EPSRC grant EP/S020527/1.

\onecolumngrid
\newpage

\setcounter{equation}{0}
\renewcommand{\theequation}{S\arabic{equation}}

\section*{SUPPLEMENTARY INFORMATION FOR ``Excitonic fractional quantum  Hall insulators in Moir\'e heterostructures''}
\begin{appendix}
\section{Electron-hole two-body problem in equal-and-opposite magnetic field}
Consider an electron and a hole with charges $-e$ and $e$ respectively, and identical masses $m$. They are confined to the 2D $x-y$ plane, and are coupled to the vector potentials $\bm{A}_e=-\bm{A}_h=\bm{A}$, where the relative sign reflects the opposite magnetic fields felt by the two particles. They interact via an attractive potential $-v_d(r)$ with $v_d(r)>0$, so that the Hamiltonian is 
\begin{equation}
H_\text{exc}=\frac{(\bm{p}_e+e\bm{A})^2}{2m}+\frac{(\bm{p}_h+e\bm{A})^2}{2m}-v_d(r).
\end{equation}

Now transform to centre-of-mass (COM) and relative coordinates 
\begin{gather}
\bm{R}=\frac{\bm{r}_h+\bm{r}_2}{e},\quad \bm{r}=\bm{r}_h-\bm{r}_e\\
\bm{P}=\bm{p}_h+\bm{p}_e,\quad \bm{p}=\frac{\bm{p}_h-\bm{p}_e}{2}.
\end{gather}
In this basis, the Hamiltonian cleanly decouples into COM and relative sectors
\begin{gather}
H_\text{exc}=H_{\bm{R}}+H_{\bm{r}}\\
H_{\bm{R}}=\frac{(\bm{P}+2e\bm{A})^2}{4m}\\
H_{\bm{r}}=\frac{(\bm{p}+\frac{e}{2}\bm{A})^2}{m}-v_d(r).
\end{gather}

It will be most convenient to use the symmetric gauge $\bm{A}=\frac{B}{2}(y\hat{x}-x\hat{y})$. The COM is a Landau level problem for particles of mass $2m$ and charge $-2e$ in a magnetic field $-B$. This should be projected to the LLL, which provides the macroscopic degeneracy of the exciton energy levels. Owing to the increased coupling $(2eB)$ to the magnetic field, the degeneracy is $2N_\Phi$, twice that of the single-particle states.  

On the other hand, the relative sector is mapped to a particle of mass $m/2$ and charge $-e/2$ in a magnetic field $-B$, subject to an attractive central potential $-v_d(r)$. Upon projection to the LLL, $H_{\bm{r}}$ becomes easy to solve since the matrix elements of the potential are diagonal in the basis of LLL angular momentum eigenstates. Therefore the binding energies are simply Haldane pseudopotentials, and are obtained using symmetric gauge wavefunctions with magnetic length $\tilde{\ell}_B=\sqrt{2}\ell_B=\sqrt{2/eB}$
\begin{equation}
E_m=-\bra{\psi_{LLL}^m}v_d(r)\ket{\psi_{LLL}^m}=-\frac{1}{2^{m-1} m! \tilde{\ell}_B^{2(m+1)}}\int dr\,v_d(r)r^{2m+1}e^{-\frac{r^2}{2\tilde{\ell}_B^2}}.
\end{equation}
For the Coulomb interaction $v_d(r)=-e^2/r$, we obtain
\begin{equation}
E_m=-\frac{e^2}{ \ell_B}\times\frac{\sqrt{\pi}}{2}\frac{(2m-1)!!}{2^m m!}.
\end{equation}

In terms of complex coordinates $z=\frac{z_h+z_e}{2}$ and $u=z_h-z_e$, the general form of an exciton eigenfunction in the $m$-th relative angular momentum channel is then 
\begin{equation}
\psi^m_{\text{exc}}\sim \tilde{f}(z)u^me^{-\frac{|z|^2}{2\ell_B^2}}e^{-\frac{|u|^2}{8\ell_B^2}}
\end{equation}
where $\tilde{f}(z)$ is analytic in $z$.

\section{Effective Two-Exciton Interaction}
For two excitons, we have two electron coordinates $z^e_1,z^e_2$ and two hole coordinates $z^h_1,z^h_2$. The interaction Hamiltonian for the four particles is
\begin{equation}\
H=v(z^e_1-z^e_2)+v(z^h_1-z^h_2)-v_d(z^e_1-z^h_1)-v_d(z^e_1-z^h_2)-v^d(z^e_2-z^h_1)-v^d(z^e_2-z^h_2)
\end{equation}
where where $v,v_d$ are the intra/intervalley interaction potentials.

In order to estimate the inter-exciton interaction, we consider the following single-exciton wavefunctions whose COM is localized at some position $\bm{R}_0$
\begin{align}
\begin{split}
\psi_{\text{exc},\bm{R}_0}(z^h,z^e)&=\frac{e^{-\frac{|R|^2}{2}}}{2\pi}e^{z(R_{0,x}-iR_{0,y})}e^{-\frac{|u|^2}{8}-\frac{|z|^2}{2}}\\
&=\frac{1}{2\pi}e^{-\frac{u^2}{8}}e^{-\frac{(\bm{R}-\bm{R}_0)^2}{2}}e^{i(R_{0,x}R_y-R_{0,y}R_x)}.
\end{split}
\end{align}
which satisfies the analyticity requirements for belonging to the lowest exciton band. We propose the following antisymmetrized state for two localized excitons separated by $\bm{R}_0$.
\begin{equation}
\Psi_{\text{exc},\bm{R}_0}(\{z^h\},\{z^e\})\sim\mathcal{A}_{e}\mathcal{A}_{h}\psi_{\text{exc},\bm{R}_0}(z^h_1,z^e_1)\psi_{\text{exc},\bm{0}}(z^h_2,z^e_2)
\end{equation}
where $\mathcal{A}_{h(e)}$ antisymmetrizes the holes (electrons). Evaluating the expectation value of the interaction, we obtain
{\small
	\begin{gather}
	V(R_0)=\frac{e^{-\frac{R_0^2}{4}}(F[I_0v]-F[I_0v_d])+e^{-\frac{3R_0^2}{4}}(F[J_0v]-F[J_0v_d])-e^{-\frac{R_0^2}{4}}F[J_0v]-e^{-\frac{3R_0^2}{4}}F[I_0v]+2e^{-\frac{R_0^2}{2}}F[v_d]}{1-2e^{-\frac{R_0^2}{2}}+e^{-R_0^2}}\\
	F[I_0V]\equiv\int_0^\infty dr\,rI_0\left(\frac{rR_0}{2}\right)e^{-\frac{r^2}{4}}V(r)\\
	F[J_0V]\equiv\int_0^\infty dr\,rJ_0\left(\frac{rR_0}{2}\right)e^{-\frac{r^2}{4}}V(r)\\
	F[V]\equiv\int_0^\infty dr\,re^{-\frac{r^2}{4}}V(r)	
	\end{gather}}
where $J_0$ and $I_0$ are the unmodified and modified Bessel functions of the first kind. The denominator reflects the normalization---as the excitons come within $\sim\ell_B$, their envelopes begin to overlap. 

\section{Exciton Projection}
Consider a LLL-projected wavefunction $\Psi(\{z^e,z^h\})$ consisting of an equal number of electrons and holes. We wish to pair up the electrons and holes, and project all their relative motions into the same exciton state $\phi(u)$. Formally this is achieved via the excitonic projection operator $\mathcal{P}_{\text{exc}}$. Let $\Psi_\sigma(\{z,u\})$ be original wavefunction written in terms of COM ($z_{i,\sigma(i)}=\frac{z^h_i+z^e_{\sigma(i)}}{2}$) and relative ($u_{i,\sigma(i)}=z^h_i-z^e_{\sigma(i)}$)  variables for a given pairing $\sigma$. Then $\mathcal{P}_{\text{exc}}$ acts as
	\begin{equation}
	\mathcal{P}_{\text{exc}}\Psi(\{z^e,z^h\})=\sum_{\sigma\in S_N}\text{sgn}\,\sigma\int d\{u'\}\Psi_\sigma(\{z,u'\})\prod_i\phi(u_{i,\sigma(i)})\phi^*(u'_{i,\sigma(i)}).
	\end{equation}

Let's apply this to the trial wavefunction proposed in the main text (Eq.~\eqref{eq:excproj}). The exciton state to be projected into is $\phi_0(u)=e^{-u^2/8}$, so the integral $\int d\{u'\}$ above is equivalent to removing terms that have any dependence on the relative coordinates (except for the LLL Gaussian factor)

{\small
\begin{align}
\begin{split}
&\Psi_{2m}(\{z^e,z^h\}) = \mathcal{P}_{\text{exc}}[\Psi_m(\{z^e\})\Psi_m(\{z^h\})]=\mathcal{P}_{\text{exc}}\prod_{i<j}(z^h_i-z^h_j)^m(z^e_i-z^e_j)^m e^{-\sum_k \frac{|z^h_k|^2}{4}-\sum_k \frac{|z^e_k|^2}{4}}\\
&=\mathcal{P}_{\text{exc}}\sum_{\sigma\in S_N}\text{sgn}\,\sigma\prod_{i<j}(z_{i,\sigma(i)}+\frac{u_{i,\sigma(i)}}{2}-z_{j,\sigma(j)}-\frac{u_{j,\sigma(j)}}{2})^m(z_{i,\sigma(i)}-\frac{u_{i,\sigma(i)}}{2}-z_{j,\sigma(j)}+\frac{u_{j,\sigma(j)}}{2})^m e^{-\sum_k \frac{|z_{k,\sigma(k)}|^2}{2}-\sum_k \frac{|u_{k,\sigma(k)}|^2}{8}}\\
&\rightarrow \sum_{\sigma\in S_N}\text{sgn}\,\sigma\prod_{i<j}(z_{i,\sigma(i)}-z_{j,\sigma(j)})^{2m}e^{-\sum_k \frac{|z_{k,\sigma(k)}|^2}{2}-\sum_k \frac{|u_{k,\sigma(k)}|^2}{8}}.
\end{split}
\end{align}}

\section{Exact Diagonalization Studies for $m=1/4$}
We performed small-scale exact diagonalization of the two-valley problem where the valleys see the same field and are at balanced filling $(\frac{1}{4},\frac{1}{4})$ and have attractive inter-valley interactions, using the DiagHam [\url{http://www.nick-ux.org/diagham/wiki/}] package for multicomponent fermions on the sphere. We consider $N_+ = N_- = 2,3,4,5$ corresponding to zero magnetization with a flux of $N_\Phi = 4,8,12,16 $ in each component. This corresponds to a shift~\cite{wenzeeshift} of $4$, which is appropriate to the excitonic trial state proposed in the main text. We take a short-range interaction written in terms of Haldane pseudopotentials as follows
\begin{equation}\label{EqnPseudopotentials}
V^1_{\tau, \tau} = -V^0_{\tau, -\tau} = E_0,\quad V^3_{\tau,\tau} = 0.4E_0
\end{equation}
where $E_0$ sets the interaction energy scale, typically $\sim e^2/\ell_B$ in the QH setting, and all other pseudopotentials set to zero. This choice stabilizes an intervalley bound state which then sees effectively  repulsive interactions. As shown in Fig.~\ref{FigsphereED}, we find a unique $L=0$ ground state for each of these fillings; since at this filling and for the chosen shift there is no  other obvious incompressible state  beyond the excitonic Laughlin trial state, we conclude that this is likely the ground state seen in these studies.

	\begin{figure}[t]
	\includegraphics[width=\linewidth,trim={0.8cm 25.0cm 2cm 0.5cm},clip=true]{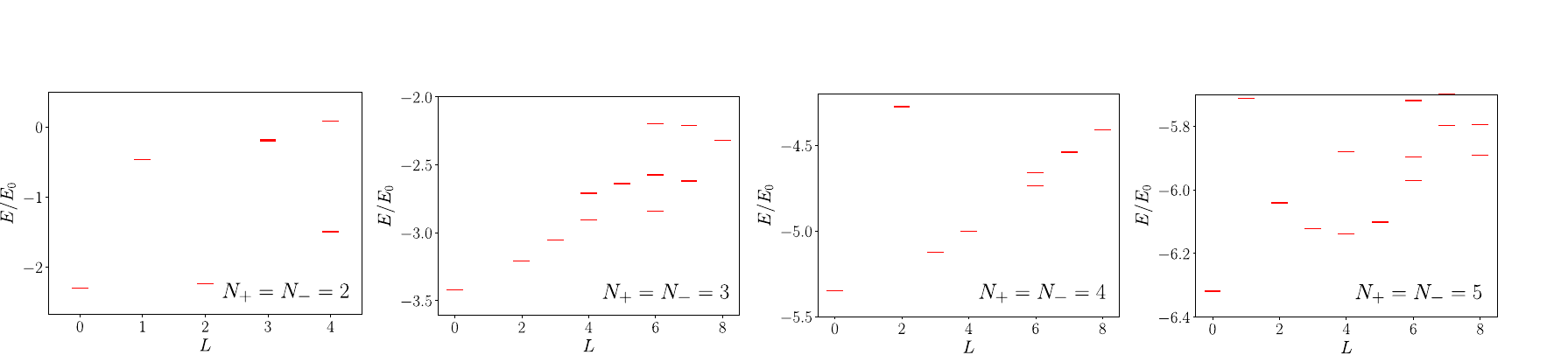}
	\caption{Exact diagonalization spectrum of low-lying states for $N_+=N_-=2,3,4,5$ fermions on a sphere interacting via Eq.~\eqref{EqnPseudopotentials}. The corresponding fluxes are $N_{\Phi}=4,8,12,16$.}
	\label{FigsphereED}
	\end{figure}

The interaction Eq.~\eqref{EqnPseudopotentials} is anisotropic in that $V_{\tau,-\tau}\neq -V_{\tau,\tau}$. We tune towards the isotropic limit through a parameter $\delta$ that controls two additional interlayer pseudopotentials $V^1_{\tau, -\tau} = -\delta E_0$ and $V^3_{\tau,-\tau} = -0.4\delta E_0$. For $\delta=1$ we obtain an `$SU(2)$-invariant' interaction $V_{\tau, -\tau}=-V_{\tau, \tau}$ (note that $V^0_{\tau,\tau}$ does not affect the spectrum). In Fig.~\ref{FigdeltaED}, we show the low-lying spectrum for $N_+=N_-=5$. For small $\delta$ there is a unique $L=0$ ground state, but there is a transition at $\delta\simeq0.6$ above which the lowest states form a $L\neq0$ multiplet. By changing $V^0_{\tau, -\tau}$ as well, we chart a preliminary phase diagram in the $V^0_{\tau, -\tau}-\delta$ plane as shown in Fig.~\ref{FigdeltaV0phasediagram}.  More extensive studies and analysis of competing phases are clearly necessary in order to fully flesh out the phase  structure.

	\begin{figure}[t]
	\includegraphics[width=\linewidth,trim={0.4cm 23.2cm 0cm 0.5cm},clip=true]{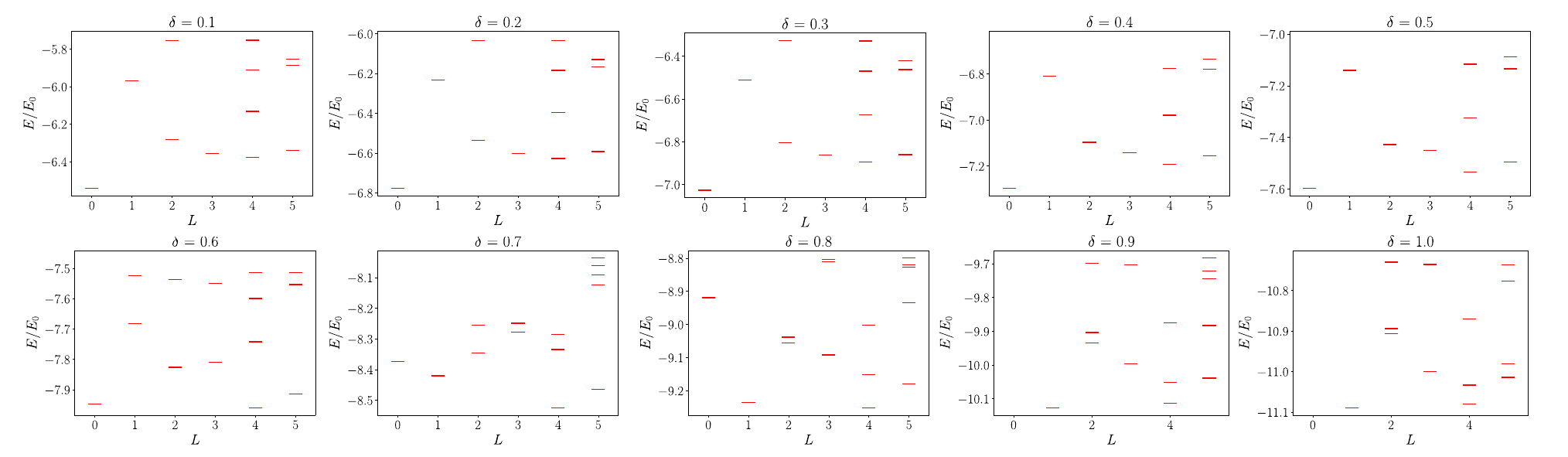}
	\caption{Exact diagonalization spectrum of low-lying states for $N_+=N_-=5$ fermions on a sphere with flux $N_\Phi=16$. The interactions are given by Eq.~\eqref{EqnPseudopotentials} augmented with $V^1_{\tau, -\tau} = -\delta E_0$ and $V^3_{\tau,-\tau} = -0.4\delta E_0$. $\delta$ is a measure of the anisotropy between intralayer and interlayer interactions, with $\delta=1$ corresponding to the $V_{\tau, -\tau}=-V_{\tau, \tau}$ limit.}
	\label{FigdeltaED}
	\end{figure}

	\begin{figure}[t]
	\includegraphics[width=0.8\linewidth,trim={0.1cm 23.5cm 6cm 0.5cm},clip=true]{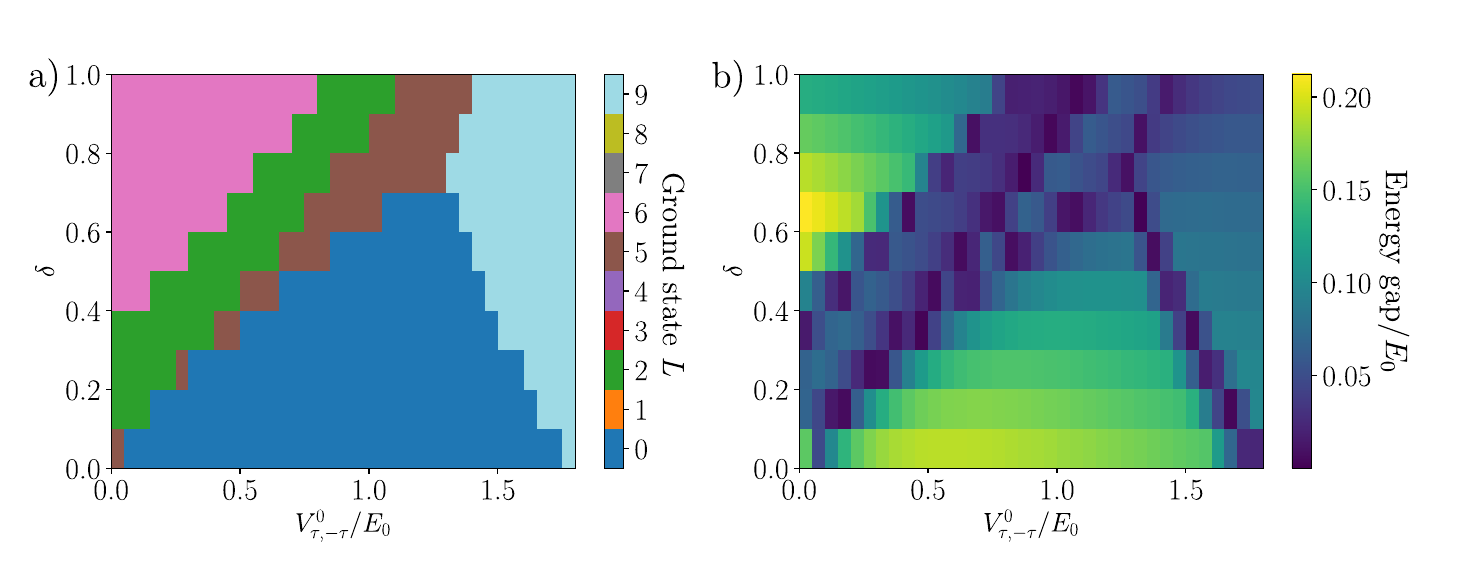}
	\caption{Phase diagram from exact diagonalization for $N_+=N_-=5$ fermions on a sphere with flux $N_\phi=16$. The interactions are Eq.~\eqref{EqnPseudopotentials} augmented with $V^1_{\tau, -\tau} = -\delta E_0$ and $V^3_{\tau,-\tau} = -0.4\delta E_0$. In addition $V^0_{\tau, -\tau}$ is tuned instead of being held fixed. a) Angular momentum $L$ of ground state multiplet. b) Energy gap between lowest and first-excited multiplets.}
	\label{FigdeltaV0phasediagram}
	\end{figure}

\end{appendix}

\end{document}